\documentclass[11pt]{article}
\usepackage[dvips]{graphicx}  
\begin{document}

\begin{center}

{\Large\bf The Michelson and Morley 1887 Experiment
and the Discovery of Absolute Motion
\rule{0pt}{13pt}}\par
\bigskip
Reginald T. Cahill \\ 
{\small\it School of Chemistry, Physics and Earth Sciences, Flinders University,
Adelaide 5001, Australia\rule{0pt}{13pt}}\\
\raisebox{-1pt}{\footnotesize E-mail: Reg.Cahill@flinders.edu.au}\\ Published: {\it Progress in Physics}
{\bf 3}, 25-29(2005).\par
\bigskip\smallskip
{\small\parbox{11cm}{%
Physics textbooks assert that in the famous interferometer 1887 experiment 
to detect absolute motion Michelson and Morley saw no rotation-induced 
fringe shifts - the signature of absolute motion; it was a null experiment.  However this is incorrect. 
Their published data revealed to them the expected fringe shifts, but 
that data gave a speed of some 8km/s using a Newtonian theory for the 
calibration of the interferometer, and so was rejected by them solely 
because it was less than the 30km/s orbital speed of the earth. A 2002
 post relativistic-effects analysis for the operation of the device 
however gives a different calibration leading to a speed $>$  300km/s. So this experiment detected both 
absolute motion and the breakdown of Newtonian physics. So far 
another six experiments have confirmed this first detection of 
absolute motion in 1887.  \rule[0pt]{0pt}{0pt}}}\bigskip
\end{center}

\section{Introduction}

The first detection of absolute motion, that is motion relative to space itself, was 
actually by Michelson and Morley in 1887 \cite{C1}. However they totally bungled the
 reporting of their own data, an achievement that Michelson managed again and again
throughout his life-long search for experimental evidence of absolute motion.  

The Michelson
interferometer was a brilliantly conceived instrument for the detection of absolute motion,
but only in 2002 \cite{C2} was its principle of operation finally understood and used to
analyse, for the first time ever, the data from the 1887 experiment, despite the enormous
impact of that experiment on the foundations of physics, particularly as they were laid down
by Einstein.  So great was Einstein's influence that the 1887 data was never re-analysed
post-1905 using a proper relativistic-effects based theory for the interferometer.  For that
reason modern-day vacuum Michelson interferometer experiments, as for example in \cite{C3},
are badly conceived, and their null results continue to cause much confusion: only a
Michelson interferometer in gas-mode can detect absolute motion, as we now see. So as better
and better vacuum interferometers were developed over the last 70 years the rotation-induced
fringe shift signature of absolute motion became smaller and smaller.  But what went
unnoticed until 2002 was that the gas in the interferometer was a key component of this
instrument when used as an `absolute motion detector', and over time the experimental
physicists were using instruments with less and less sensitivity; and in recent years they
had finally perfected a totally dud instrument.  Reports from such experiments  claim  
 that absolute motion
is not observable, as Einstein had postulated, despite the fact that the apparatus is
totally insensitive to absolute motion.    It must be emphasised that absolute motion is not
inconsistent with the various well-established relativistic effects; indeed the evidence is
that absolute motion is the cause of these relativistic effects, a proposal that goes back
to Lorentz in the 19th century.  Then of course one must use a relativistic theory for the
operation of the Michelson interferometer.   What also follows from these experiments is
that the Einstein-Minkowski spacetime ontology is invalidated, and in particular that
Einstein's postulates regarding the invariant speed of light have always been in
disagreement with experiment from the beginning.  This does not imply that the use of a
mathematical spacetime is not permitted; in quantum field theory the mathematical spacetime
encodes absolute motion effects. An ongoing confusion in physics is that absolute motion is
incompatible with Lorentz symmetry, when the evidence is that it is the cause of that
dynamical symmetry.

\section{Michelson Interferometer}
The Michelson interferometer compares the change in the difference between 
travel times, when the device is rotated, for two coherent beams of light that travel in
orthogonal directions between mirrors; the  changing time difference being indicated by the
shift of the interference fringes during the rotation.  This effect is caused by the
absolute motion of the device through 3-space with speed $v$, and that the speed of light is
relative to that 3-space, and not relative to the apparatus/observer. However to detect the
speed of the apparatus through that 3-space gas must be present in the light paths for
purely technical reasons. A theory is required to calibrate this device, and it turns
 out that the calibration of gas-mode Michelson
interferometers was only worked out in 2002. The post relativistic-effects theory for 
this device is remarkably simple.  The Fitzgerald-Lorentz
contraction effect causes the arm AB parallel to the absolute velocity to be 
physically contracted to length
\begin{equation}
L_{||}=L\sqrt{1-\frac{v^2}{c^2}}.
\label{eqn:e1}\end{equation}

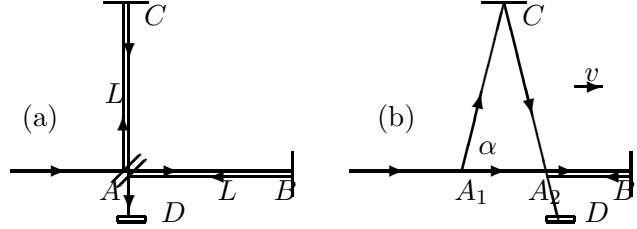
\begin{figure}
\setlength{\unitlength}{0.75mm}
\hspace{35mm}\begin{picture}(0,30)
\thicklines
\put(-10,0){\line(1,0){50}}
\put(-5,0){\vector(1,0){5}}
\put(40,-1){\line(-1,0){29.2}}
\put(15,0){\vector(1,0){5}}
\put(30,-1){\vector(-1,0){5}}
\put(10,0){\line(0,1){30}}
\put(10,5){\vector(0,1){5}}
\put(11,25){\vector(0,-1){5}}
\put(11,30){\line(0,-1){38}}
\put(11,-2){\vector(0,-1){5}}
\put(8.0,-2){\line(1,1){5}}
\put(9.0,-2.9){\line(1,1){5}}
\put(6.5,30){\line(1,0){8}}
\put(40,-4.5){\line(0,1){8}}
\put(5,12){ $L$}
\put(4,-5){ $A$}
\put(35,-5){ $B$}
\put(25,-5){ $L$}
\put(12,26){ $C$}
\put(9,-8){\line(1,0){5}}
\put(9,-9){\line(1,0){5}}
\put(14,-9){\line(0,1){1}}
\put(9,-9){\line(0,1){1}}
\put(15,-9){ $D$}
\put(50,0){\line(1,0){50}}
\put(55,0){\vector(1,0){5}}
\put(73,0){\vector(1,0){5}}
\put(85,0){\vector(1,0){5}}
\put(90,15){\vector(1,0){5}}
\put(100,-1){\vector(-1,0){5}}
\put(100,-4.5){\line(0,1){8}}
\put(68.5,-1.5){\line(1,1){4}}
\put(69.3,-2.0){\line(1,1){4}}
\put(70,0){\line(1,4){7.5}}
\put(70,0){\vector(1,4){3.5}}
\put(77.5,30){\line(1,-4){9.63}}
\put(77.5,30){\vector(1,-4){5}}
\put(73.5,30){\line(1,0){8}}
\put(83.3,-1.5){\line(1,1){4}}
\put(84.0,-2.0){\line(1,1){4}}
\put(100,-1){\line(-1,0){14.9}}
\put(73,3){$\alpha$}
\put(67,-5){ $A_1$}
\put(80,-5){ $A_2$}
\put(85,-8){\line(1,0){5}}
\put(85,-9){\line(1,0){5}}
\put(90,-9){\line(0,1){1}}
\put(85,-9){\line(0,1){1}}
\put(90,-9){ $D$}
\put(95,-5){ $B$}
\put(79,26){ $C$}
\put(90,16){ $v$}
\put(-8,8){(a)}
\put(55,8){(b)}

\end{picture}

\vspace{10mm}
\caption{\small{Schematic diagrams of the Michelson Interferometer, with beamsplitter/mirror at $A$ and
mirrors at $B$ and $C$ on arms  from $A$, with the arms of equal length $L$ when at rest.  $D$ is the
detector screen. In (a) the interferometer is
at rest in space. In (b) the interferometer is moving with speed $v$ relative to space in the direction
indicated. Interference fringes are observed at  $D$.  If the interferometer is
rotated in the plane  through $90^o$, the roles of arms $AC$ and $AB$ are interchanged, and during the
rotation shifts of the fringes are seen in the case of absolute motion, but only if the apparatus operates
in a gas.  By measuring fringe shifts the speed $v$ may be determined.}  \label{fig:Minterferometer}}
\end{figure}

The time $t_{AB}$ to travel $AB$ is set by $Vt_{AB}=L_{||}+vt_{AB}$ , while for $BA$ by 
$Vt_{BA}=L_{||}-vt_{BA}$, where $V=c/n$   is the speed of light,  with
$n$  the refractive index of the gas present (we ignore here the Fresnel
drag effect for simplicity - an effect caused by the gas also being in absolute motion).
For the total ABA travel time we then obtain
\begin{equation}
t_{ABA}=t_{AB}+t_{BA}=\frac{2LV}{V^2-v^2}\sqrt{1-\frac{v^2}{c^2}}.
\label{eqn:e2}\end{equation} 
For travel in the $AC$ direction we have, from the \newline Pythagoras theorem for the
 right-angled triangle in Fig.1 that $(Vt_{AC})^2=L^2+(vt_{AC})^2$  and that
$t_{CA}=t_{AC}$. Then for the total $ACA$ travel time
\begin{equation}
t_{ACA}=t_{AC}+t_{CA}=\frac{2L}{\sqrt{V^2-v^2}}.
\label{eqn:e3}\end{equation} 
 Then the difference in travel time is
\begin{equation}
\Delta t=\frac{(n^2-1)L}{c}\frac{v^2}{c^2}+O\left(\frac{v^4}{c^4} \right).
\label{eqn:e4}\end{equation}                                                                                  
after expanding in powers of $v/c$. This clearly shows that the interferometer 
can only operate as a detector of absolute motion when not in vacuum ($n=1$), namely when
the light passes through a gas, as in the early experiments (in transparent solids a more
complex phenomenon occurs and rotation-induced fringe shifts from absolute motion do not
occur). A more general analysis \cite{C2,C9,C10}, including Fresnel drag, gives
\begin{equation}
\Delta t=k^2\frac{Lv_P^2}{c^3}\cos(2(\theta-\psi)).
\label{eqn:e5}\end{equation}                                                                 
where $k^2\approx n(n^2-1)$, while neglect of the Fitzgerald-Lorentz contraction effect
gives   $k^2\approx n^3 \approx 1$
 for gases, which is essentially the Newtonian calibration that \newline Michelson used.  All the
rotation-induced fringe shift data from the 1887 Michelson-Morley experiment, as tabulated
in \cite{C1}, is shown in Fig.2.  The existence of this data continues to be denied by the
world of physics.

\begin{figure}
\hspace{30mm}\includegraphics[scale=1.1]{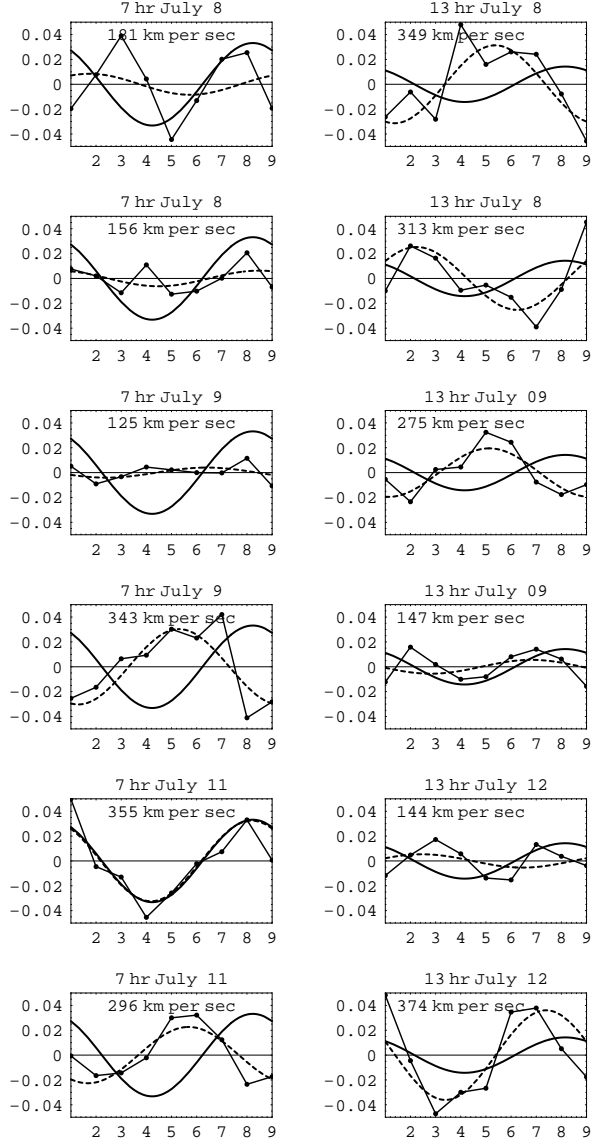}
\caption{{\small Shows all the Michelson-Morley 1887 data after removal of the
temperature induced fringe drifts.  The data for each $360^0$ full turn (the average of 6 individual turns) is
divided into the 1st and 2nd $180^0$ parts and  plotted one above the other.  The dotted curve
shows a best fit to the data using (\ref{eqn:e5}), while the full curves show the expected forms using the
Miller direction for ${\bf v}$ and the location and times of the Michelson-Morley observations. While the amplitudes
are in agreement in general with the Miller based predictions, the phase varies somewhat. This may be related 
to the Hick's effect \cite{C4}
when, necessarily, the mirrors are not orthogonal.  We see that this data corresponds to a speed in excess 
of 300km/s, and not the 8km/s reported
in
\cite{C1}, which was based on using Newtonian physics to calibrate the interferometer. }\label{fig:MMplots}}
\end{figure}

\begin{figure}
\hspace{3mm}\includegraphics[scale=0.8]{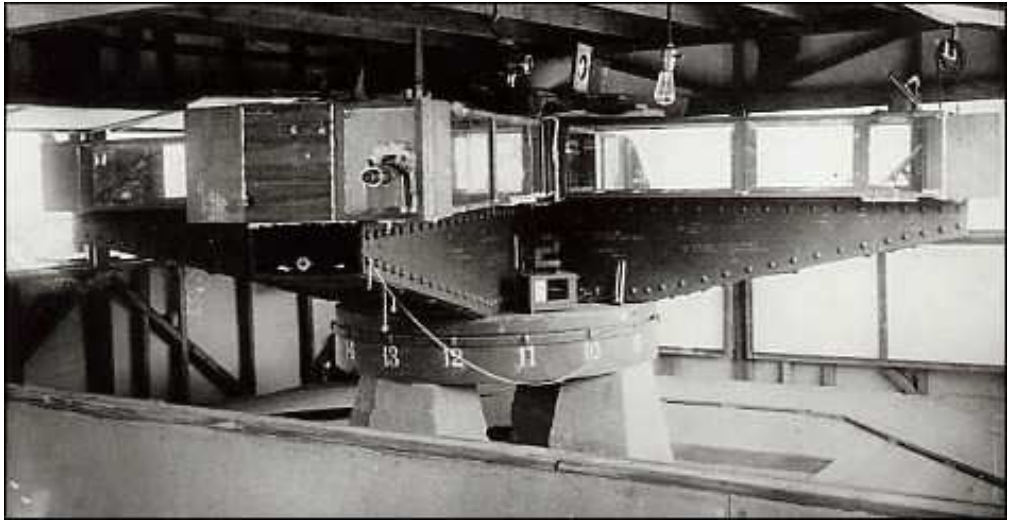}
\caption{\small{ Miller's interferometer with an effective arm length of $L=32$m achieved by multiple
reflections. Used by Miller on Mt.Wilson to perform the
1925-1926 observations of absolute motion. The steel arms weighed 1200 kilograms and floated in a tank of
275 kilograms of Mercury. From Case Western Reserve University Archives.}  
\label{fig:Millerinterf}}\end{figure}
 
The interferometers are operated with the arms horizontal, as shown by Miller's 
interferometer in Fig.3. Then in (\ref{eqn:e5}) $\theta$ is the azimuth of one arm
(relative to the local meridian), while $\psi$ is the azimuth of the absolute motion
velocity projected onto the plane of the interferometer, with projected component $v_P$.
Here the Fitzgerald-Lorentz contraction is a real dynamical effect of absolute motion,
unlike the Einstein spacetime view that it is merely a spacetime perspective artefact, and
whose magnitude depends on the choice of observer. The instrument is operated by rotating
at a rate of one rotation over several minutes, and observing the shift in the fringe
pattern through a telescope during the rotation.  Then fringe shifts from six (Michelson
and Morley) or twenty (Miller) successive rotations are averaged, and the average sidereal
time noted, giving in the case of Michelson and Morley the data in Fig.2, or the Miller
data like that in Fig.4. The form in (\ref{eqn:e5}) is then fitted to such data, by varying
the parameters $v_P$ and  $\psi$.  However Michelson and Morley implicitly assumed the
Newtonian value  $k=1$, while Miller used an indirect method to estimate the value of $k$,
as he understood that the Newtonian theory was invalid, but had no other theory for the
interferometer.  Of course the Einstein postulates have that absolute motion has no
meaning, and so effectively demands that $k=0$.  Using $k=1$  gives only a nominal value
for $v_P$, being some 8km/s for the Michelson and Morley experiment, and some 10km/s from
Miller; the difference arising from the different latitude of Cleveland and Mt. Wilson. 
The relativistic theory for the calibration of gas-mode interferometers was first used in
2002,\cite{C2}.

\section{Michelson-Morley Data}
Fig.2 shows all the Michelson and Morley air-mode interferometer fringe shift data, 
based upon a total of only 36 rotations in July 1887, revealing the nominal speed of some
8km/s when analysed using the prevailing but incorrect Newtonian theory which has  $k=1$ in 
(\ref{eqn:e5}); and this value was known to Michelson and Morley. Including the Fitzgerald-Lorentz
dynamical contraction effect as well as the effect of the gas present as in (\ref{eqn:e5})  we find that
$n_{air}=1.00029$ gives $k^2=0.00058$  for air, which explains why the observed fringe shifts were so
small. We then obtain the speeds shown in Fig.2. In some cases the data does not have the expected form
in  (\ref{eqn:e5}); because the device was being operated at almost the limit of sensitivity. The
remaining fits give a speed in excess of 300km/s.   The often-repeated statement that Michelson and Morley did not see any
rotation-induced fringe shifts is completely wrong; all physicists should read their paper
\cite{C1} for a re-education, and indeed their paper has a table of the observed fringe
shifts. To get the Michelson-Morley Newtonian based value of some 8km/s we must multiply the
above speeds by $k=\sqrt{0.00058}=0.0241$. They rejected their own data on the sole but spurious ground
that the value of 8km/s was smaller than the speed of the earth about the sun of 30km/s.  What their
result really showed was that (i) absolute motion had been detected because fringe shifts of
the correct form, as in  (\ref{eqn:e5}), had been detected, and (ii) that the theory giving $k^2=1$  was
wrong, that Newtonian physics had failed. Michelson and Morley in 1887 should have announced that
the speed of light did depend of the direction of travel, that the speed  was relative to an
actual physical 3-space. However contrary to their own data they concluded that absolute
motion had not been detected.  This bungle has had enormous implications for fundamental
theories of space and time over the last 100years, and the resulting confusion is only now
being finally corrected.

\begin{figure}[t]
\hspace{30mm}\includegraphics[scale=1.12]{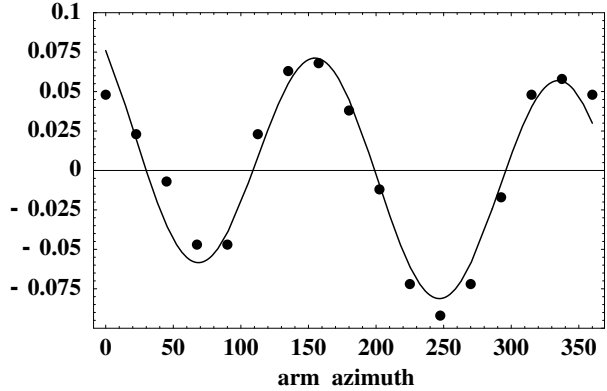}
\caption{\small Typical Miller rotation-induced fringe shifts from average of 20 rotations, 
measured every 22.5$^0$, in fractions of a wavelength $\Delta \lambda/\lambda$ , vs  azimuth $\theta$  (deg), measured
clockwise from North, from Cleveland Sept. 29, 1929 16:24 UT; 11:29 average sidereal time. This shows the quality of
the fringe data that Miller obtained, and is considerably better than the comparable data by Michelson and Morley in
Fig.2. The curve is the best fit using the form in (\ref{eqn:e5}) but including a Hick's \cite{C4} $\cos(\theta-\beta)$ component
that is required when the mirrors are not orthogonal, and gives
$\psi=158^0$ , or
$22^0$  measured from South, and a projected speed of $v_P =351$ km/s.  This value for $v$ is different from that in Fig.2
because of the difference in latitude of Cleveland and Mt. Wilson. This process was repeated some 12,000 times over
days and months throughout 1925/1926 giving, in part, the data in Fig.5. }  
\label{fig:Miller}\end{figure}

\begin{figure}[t]
\hspace{15mm}\includegraphics[scale=1.5]{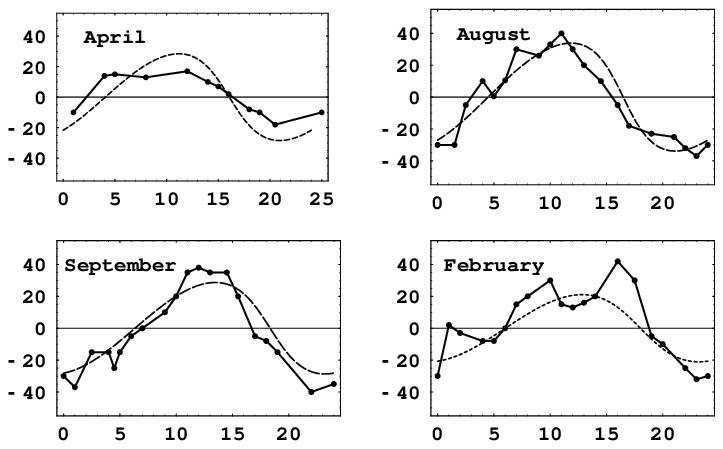}
\caption{\small  Miller azimuths $\psi$, measured from south and plotted against sidereal time in hrs, showing both
data and best fit of theory giving
$v=433$ km/s in the direction ($\alpha=5.2^{hr}, \delta=-67^0$), using $n=1.000226$ appropriate for the
altitude of Mt. Wilson. The variation form month to month arises from the orbital motion of the earth about the sun: in different
months the vector sum of the galactic velocity of the solar system with the orbital velocity and sun in-flow velocity is different.  As shown
in Fig.\ref{fig:DeWittetimes} DeWitte using a completely different experiment detected the same direction and speed.}  
\label{fig:MillerAz}\end{figure}

\section{Miller Interferometer}
It was Miller \cite{C4} who saw the flaw in the 1887 paper and realised that the theory for
the Michelson interferometer must be wrong.  To avoid using that theory Miller introduced
the scaling factor $k$, even though he had no theory for its value. He then used the effect of
the changing vector addition of the earth's orbital velocity and the absolute galactic
velocity of the solar system to determine the numerical value of  $k$, because the orbital
motion modulated the data, as shown in Fig.5. By making some 12,000 rotations of the
interferometer at Mt. Wilson in 1925/26 Miller determined the first estimate for $k$  and for
the absolute linear velocity of the solar system. Fig.4 shows typical data from averaging
the fringe shifts from 20 rotations of the Miller interferometer, performed over a short
period of time, and clearly shows the expected form in (\ref{eqn:e5}) (only a linear drift caused by
temperature effects on the arm lengths has been removed - an effect also removed by
Michelson and Morley and also by Miller). In Fig.4 the fringe shifts during rotation are
given as fractions of a wavelength, $\Delta \lambda/\lambda=\Delta t/T$, where $\Delta t$ is given by 
(\ref{eqn:e5}) and $T$  is the period of the light. Such rotation-induced fringe shifts clearly show that
the speed of light is different in different directions. The claim that Michelson interferometers,
operating in gas-mode, do not produce fringe shifts under rotation is clearly incorrect. But it is that
claim that lead to the continuing belief, within physics, that absolute motion had never been detected,
and that the speed of light is invariant. The value of $\psi$  from such rotations together lead
to plots like those in Fig.5, which show $\psi$ from the 1925/1926 Miller \cite{C4}
interferometer data for four different months of the year, from which the RA = 5.2hr is
readily apparent. While the orbital motion of the earth about the sun slightly affects the
RA in each month, and Miller used this effect do determine the value of  $k$, the new theory of
gravity required a reanalysis of the data \cite{C9,C11}, revealing that the solar system has
a large observed galactic velocity of some 420$\pm$30km/s in the direction (RA=5.2hr, Dec=
-67deg). This is different from the speed of 369km/s in the direction (RA=11.20hr, Dec=
-7.22deg) extracted from the Cosmic Microwave Background (CMB) anisotropy, and which
describes a motion relative to the distant universe, but not relative to the local 3-space
(The Miller velocity is explained by galactic gravitational in-flows; see \cite{C12}.)

    Two old interferometer experiments, by Illingworth \cite{C5} and Joos \cite{C6}, used
helium, enabling the refractive index effect to be recently confirmed, because for helium,
with $n=1.000036$, we find that $k^2=0.00007$.  Until the refractive index effect was taken into account
the data from the helium-mode experiments appeared to be inconsistent with the data from the air-mode
experiments; now they are seen to be consistent. Ironically helium was introduced in place of air to
reduce any possible unwanted effects of a gas, but we now understand the essential role of
the gas. The data from an interferometer experiment by Jaseja {\it  et al} \cite{C7}, using two
orthogonal masers with a He-Ne gas mixture, also indicates that they detected absolute
motion, but were not aware of that as they used the incorrect Newtonian theory and so
considered the fringe shifts to be too small to be real, reminiscent of the same mistake by
Michelson and Morley. The Michelson interferometer is a 2nd order device, as the effect of
absolute motion is proportional to  $(v/c)^2$, as in  (\ref{eqn:e5}).

\section{1st Order Experiments}
However much more sensitive 1st order  experiments are also possible.  Ideally they simply 
measure the change in the one-way EM travel-time as the direction of propagation is changed.  Fig.6
 shows the North-South orientated coaxial 
cable Radio Frequency (RF) travel time variations measured by DeWitte in Brussels in 1991,
\cite{C9,C10,C11}, which gives the same RA of absolute motion as found by Miller. That
experiment showed that RF waves travel at speeds determined by the orientation of the cable
relative to the Miller direction. That these very different experiments show the same speed
and RA of absolute motion is one of the most startling discoveries of the twentieth century.
Torr and Kolen \cite{C8} using an East-West orientated nitrogen gas-filled coaxial cable also
detected absolute motion.  It should be noted that analogous optical fibre experiments give
null results for the same reason, apparently, that transparent solids in a Michelson
interferometer also give null results, and so behave differently to coaxial cables.

\begin{figure}[t]
\hspace{25mm}\includegraphics[scale=1.2]{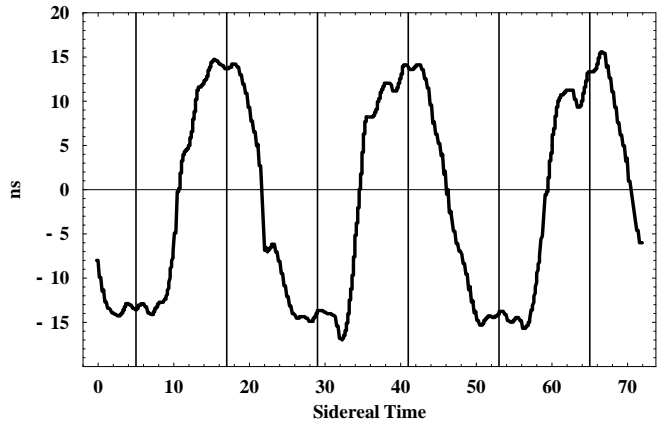}
\caption{\small{ Variations in twice the one-way travel time, in ns, for an RF signal to travel 1.5
km through a coaxial cable between  Rue du Marais and Rue de la Paille, Brussels. 
An offset  has been used  such that the average is zero.   The cable has a
North-South  orientation, and the data is the difference of the travel times  for NS and SN
propagation.  The sidereal time for maximum  effect of $\sim\!\!5$hr and   $\sim\!\!17$hr (indicated
by vertical lines) agrees with the direction found by Miller. Plot shows
data over 3 sidereal days  and is plotted against sidereal time. DeWitte recorded such data from 178 days, and confirmed that the effect tracked
sidereal time, and not solar time. Miller also confirmed this sidereal time tracking.
 The fluctuations are evidence of turbulence in
  the flow.}  
\label{fig:DeWittetimes}}\end{figure}

Modern resonant-cavity interferometer experiments, for which the analysis leading to  (\ref{eqn:e5})
 is applicable, use vacuum with $n=1$, and then $k=0$, predicting no rotation-induced fringe
shifts.  In analysing the data from these experiments the consequent null effect is
misinterpreted, as in
\cite{C3}, to imply the absence of absolute motion. But it is absolute motion which
 causes the dynamical effects of
length contractions, time dilations and other relativistic effects, in accord with 
Lorentzian interpretation of
relativistic effects.  The detection of absolute motion is not incompatible with 
Lorentz symmetry; the contrary
belief was postulated by Einstein, and has persisted for over 100 years, since 1905.
 So far the experimental
evidence is that absolute motion and Lorentz symmetry are real and valid phenomena;
absolute motion is motion
presumably relative to some substructure to space, whereas Lorentz symmetry
 parameterises dynamical effects caused
by the motion of systems through that substructure. To check Lorentz symmetry 
we can use vacuum-mode resonant-cavity
interferometers, but using gas within the resonant-cavities would enable 
these devices to detect absolute motion
with great precision. As well there are novel wave phenomena that could also 
be studied, see \cite{C9,C10}.

\section{Conclusions}
So absolute motion was first detected in 1887, and again in at least another 
six experiments over the last 100 years.  Had Michelson and Morley been as 
astute as their
younger colleague Miller, and had been more careful in reporting their {\it non-null} data, the
history of physics over the last 100 years would have totally different, and the spacetime
ontology would never have been introduced. That ontology was only mandated by the mistaken
belief that absolute motion had not been detected.  By the time Miller had sorted out that
bungle, the world of physics had adopted the spacetime ontology as a model of reality
because that model appeared to be confirmed by many relativistic phenomena, mainly from
particle physics, although these phenomena could equally well have been understood using
the Lorentzian interpretation which involved no spacetime.  We should now understand that
in  quantum field theory a mathematical spacetime encodes absolute motion effects upon the
elementary particle systems, but that there exists a physically observable foliation of
that spacetime into a geometrical model of time and a separate geometrical model of
3-space.





\end{document}